\documentclass[conference]{IEEEtran}
\IEEEoverridecommandlockouts
\usepackage{algorithmicx}
\usepackage{cite}
\usepackage{amsmath,amssymb,amsfonts}
\usepackage{algorithm}
\usepackage{algorithmicx}
\usepackage{algpseudocode}  
\usepackage{graphicx}
\usepackage{booktabs}
\usepackage{textcomp}
\usepackage{xcolor}
\usepackage{algorithm}
\usepackage[colorlinks=true, linkcolor=blue, citecolor=blue, urlcolor=blue]{hyperref}
\usepackage{hyperref}
\usepackage{cleveref}
\usepackage{subfig}  
\usepackage{titlesec}
\usepackage{indentfirst}

\titlespacing*{\section}{3pt}{3pt}{3pt}
\titlespacing*{\subsection}{3pt}{3pt}{3pt}
\def\BibTeX{{\rm B\kern-.05em{\sc i\kern-.025em b}\kern-.08em
	T\kern-.1667em\lower.7ex\hbox{E}\kern-.125emX}}
\begin{document}

\title{Relative Localization of UAV Swarms in GNSS-Denied Conditions
}

\author{ Guangyu Lei$^* $$^\dagger$, Yuqi Ping$^*$$^\dagger$, Tianhao Liang$^*$$^\dagger$, Huahao Ding$^*$$^\dagger$, Tingting Zhang$^*$$^\dagger$$^\ddagger$\\
	$^*$School of Information Science and Technology, Harbin Institute of Technology, Shenzhen, P. R. China\\
	$^\dagger$Guangdong Provincial Key Laboratory of Space-Aerial Networking and Intelligent Sensing, Shenzhen, P. R. China\\
	$^\ddagger$Peng Cheng Laboratory, Shenzhen, P. R. China\\
	Contact Email: zhangtt@hit.edu.cn

\thanks{This paper was supported by the NSFC under Grant No. 62501194 and 62171160, the Guangdong Provincial Key Laboratory (2024) (No.2024KSYS023), and KJZD20240903100022029, and also Major Key Project of PCL (PCL2024A01).}}

\maketitle

\begin{abstract}
Relative localization of unmanned aerial vehicle (UAV) swarms in global navigation satellite system (GNSS) denied environments is essential for emergency rescue and battlefield reconnaissance. Existing methods suffer from significant localization errors among UAVs due to packet loss and high computational complexity in large swarms. This paper proposes a clustering-based framework where the UAVs simultaneously use communication signals for channel estimation and ranging. Firstly, the spectral clustering is utilized to divide the UAV swarm into different sub-clusters, where matrix completion and multidimensional scaling yield high-precision relative coordinates. Subsequently, a global map is created by the inter-cluster anchor fusion. A case study of UAV integrated communication and sensing (ISAC) system is presented, where the Orthogonal Time Frequency Space (OTFS) is adopted for ranging and communication. Experimental results show that the proposed method reduces localization errors in large swarms and loss of range information. It also explores the impact of signal parameters on communication and localization, highlighting the interplay between communication and localization performance.
\end{abstract}
\vspace{-10pt}

\section{Introduction}

Unmanned aerial vehicle (UAV) swarm systems, with advancements in wireless communication and autonomous control, offer significant benefits in firefighting, rescue, and terrain exploration\cite{rescue, UAV2, UAV3,Jiacheng,Jiacheng2}. However, challenges arise in emergency response, battlefield reconnaissance, and high-altitude sensing due to the absence of global navigation satellite system (GNSS) signals \cite{UAV4,UAV5,swarm, Denied,YUAN2}. Traditional positioning based on address broadcasting faces two significant issues: dependence on GNSS for timing and coordinates, and susceptibility to interference and channel contention in large swarms due to the lack of media access control in a single-channel broadcast system, respectively \cite{datalink}.

Recent research focused on relative localization methods using onboard sensors, reconstructing position through inter-UAV range and angle measurements. Multidimensional scaling (MDS), a classical technique in wireless sensor networks, maps range matrices to low-dimensional coordinates and has been extended to UAV swarm localization \cite{borg2007modern}. However, conventional MDS suffers from two main challenges: incomplete range matrices due to missing measurements, leading to localization failure, and the exponential computational burden of centralized architectures as swarm size grows, imposing a significant burden on real-time systems. To address these issues, MDS-MAP employs minimum-hop estimates to compensate for missing range information \cite{shang2003localization}. Additionally, SMDS(P) incorporates angle-of-arrival information to enhance localization accuracy \cite{swarm}. The coalition game-based clustering strategies decompose UAV swarm into intra-cluster localization and inter-cluster map merging \cite{Denied}. Notably, the authors in \cite{OTFSJour} introduce Orthogonal Time Frequency Space (OTFS) modulation, utilizing delay-Doppler orthogonality for joint communication and sensing in Line-of-Sight (LoS) and Non-Line-of-Sight conditions, offering a lightweight localization solution without dedicated ranging modules and a new ISAC paradigm for UAV swarm localization\cite{YUAN}.

Despite these advancements, several challenges remain: conventional clustering strategies rely on iterative measurements or complex theoretical processes, introducing additional communication overhead and computational latency\cite{swarm}\cite{Denied}. The ISAC framework has not thoroughly explored how physical-layer signal parameters influence localization accuracy, and the assumption of a fully connected network overlooks realistic communication range constraints, leading to an underestimation of long-range ranging failures\cite{OTFSJour}.

This paper presents a cooperative localization framework using spectral clustering, with the main contributions summarized as follows:
\begin{enumerate}
	\item We propose a spectral clustering strategy that converts inter-UAV channel estimation results into a similarity graph for clustering. UAVs are grouped based on proximity, followed by matrix completion and MDS for relative localization, significantly improving accuracy in large swarms and with range information loss.

	\item A case study of OTFS signals, suitable for high-dynamic environments, is introduced in the paper, which successfully employs communication signals for relative localization of the UAV swarm, constructing an ISAC system for UAV swarm relative localization.

	\item Simulation results are executed to examine how the communication signal structure affects localization, emphasizing the dual impact of communication signal parameters on localization and communication performance.
%
%
%
%
\end{enumerate}

\begin{figure*}[htbp]
	\centering
	\includegraphics[scale=0.105]{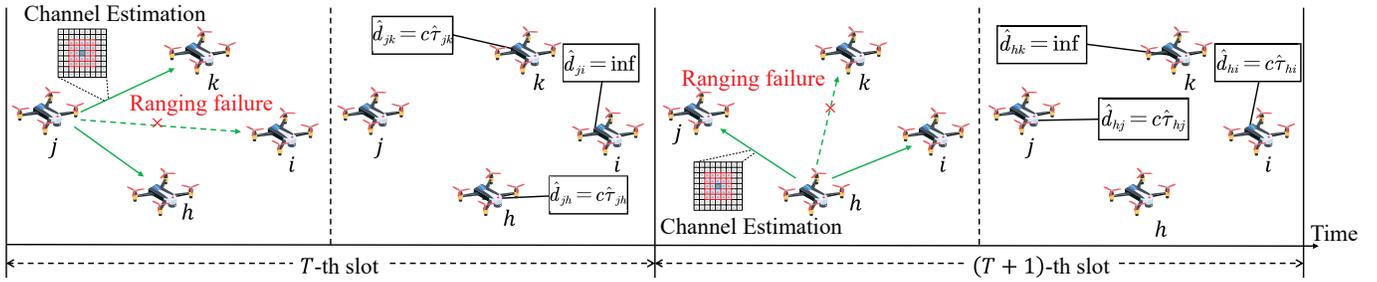} 
\vspace{-20pt}
	\caption{UAV network ranging process and rules.}
	\label{fig:Agree}
\vspace{-20pt} 
\end{figure*}
\section{System Model}

Consider a 3D space when there are a total of $L$ UAVs, the 3D position of the UAV swarm can be written as $\mathbf{P}=[\mathbf{p}_{1},\mathbf{p}_{2},\cdots,\mathbf{p}_{L}]^{\top}\in\mathbb{R}_{+}^{L\times 3}$. For the range $d_{ij}$ between UAV $i$ and UAV $j$ can be expressed as
\begin{equation}
d_{ij}=\sqrt{||\mathbf{p}_{i}-\mathbf{p}_{j}||^{2}_{2}}.
\end{equation}
%
%

The system uses a centrally scheduled polling-based channel estimation protocol to construct a global delay matrix for the UAV swarm. An edge server or centralized controller designates each UAV as the transmitter according to a predefined schedule, while the remaining UAVs serve as receivers, performing channel estimation. 

Due to spatial constraints and limited connectivity, not all UAVs may receive the transmitted signal. For those that fail to detect the signal, the range to the transmitter is set to infinity to indicate an unreachable link explicitly
\begin{equation}\label{dis}
\hat{d}_{ij}=\begin{cases}
c\hat{\tau}_{ij}&	\operatorname{Detect\ sucessufully}\\
\operatorname{inf}&	\operatorname{Fail\ to\ detetct}\\
\end{cases},
\end{equation}
where $\hat{\tau_{ij}}$ is delay results from channel estimation when UAV $i$ communicates with UAV $j$, as shown in Fig.~\ref{fig:Agree}. 

After completing channel estimation, each UAV sends its results to a relay node with a stable communication link. The relay node aggregates data from other UAVs in the cluster and uploads it to the centralized node or edge server. The edge server receives and processes the aggregated data, compiling it into a global three-dimensional range estimation matrix $ \mathbf{D}$ by \eqref{dis}.

\begin{equation}\label{Dmat}
\mathbf{D} = \left( \begin{matrix} 
0&\hat{d}_{1,2} & \cdots & \hat{d}_{1,L} \\ 
\hat{d}_{2,1}& 0&\cdots & \hat{d}_{2,L}\\
\vdots & \vdots &\ddots & \vdots \\ 
\hat{d}_{L,1}&\hat{d}_{L,2} & \cdots & 0 \\ 
\end{matrix} \right), 
\vspace{-8pt}
\end{equation}
where $ \hat{d}_{ij} $ represents the estimation of range derived from the delay extracted from the signal received by UAV $ i $ from UAV $ j $, i.e., the range between UAV $ i $ and UAV $ j $ obtained through channel estimation. All subsequent network localization and computation tasks are handled at the centralized node, ensuring efficient processing and control of the swarm's localization \cite{OTFSJour}.

\section{Proposed Algorithm}
\subsection{UAV Clustering}
\subsubsection{Spectral Clustering}

Based on the obtained inter-UAV range list, clustering relies solely on the ranges between UAVs. Spectral clustering is introduced to address this, with UAV similarity as the basis for clustering \cite{spect}. A similarity graph is first constructed, and clustering is then performed using the graph's spectral properties (i.e., eigenvectors). To map the UAV ranges to node similarities, a Gaussian kernel function is introduced, defining the similarity matrix as $ \mathbf{W} $, the elements of the matrix $W_{ij}$ can be obtained by \eqref{Dmat}
\begin{equation}
W_{ij} =\begin{cases} 
\exp \left( -\frac{\hat{d}_{ij}^2}{2\sigma^2} \right) &\hat{d}_{ij}<\text{inf} \\ 
0 &\hat{d}_{ij}=\text{inf}\\
\end{cases} ,
\end{equation}
where $\sigma$ is the Gaussian kernel bandwidth parameter, often chosen based on the range information distribution, which is typically set to the average or median range between UAVs in the swarm. The degree matrix $\mathbf{\Gamma}$ and the Laplacian matrix $\mathbf{L}$ can be constructed from $\mathbf{W}$. $\mathbf{\Gamma}$ is a diagonal matrix where $\Gamma_{i,i}$ represents the degree of node $i$, computed as
\begin{equation}
{\Gamma}_{i,i} = \sum_{j} W_{ij}.
\end{equation}
The normalized Laplacian matrix $ \mathbf{L} $ is defined as
\begin{equation}
\mathbf{L} = \mathbf{\Gamma}^{-\frac{1}{2}} \left( \mathbf{D} - \mathbf{W} \right) \mathbf{\Gamma}^{-\frac{1}{2}},
\end{equation}
where $ \mathbf{\Gamma} \in \mathbb{R}^{L \times L} $ is the diagonal matrix, with $ \mathbf{\Gamma}^{-\frac{1}{2}} $ its inverse square root. Eigenvalue decomposition of the normalized Laplacian matrix $ \mathbf{L} $ yields eigenvectors $ \mathbf{U} = \{\mathbf{v}_1, \mathbf{v}_2, \dots, \mathbf{v}_L\} $ and eigenvalues $ \boldsymbol{\varLambda} = \operatorname{diag}\{\lambda_1, \lambda_2, \dots, \lambda_L\} $. The eigenvalue distribution of $ \mathbf{L} $ reveals the data’s clustering structure in spectral clustering. The eigenvalues converge at the $ k $-th eigenvalue, implying the whole graph can be divided into $ k $ connected subgraph. The first $ k $ eigenvectors form the matrix $ \mathbf{U}_k = [\mathbf{v}_1, \mathbf{v}_2, \dots, \mathbf{v}_k] $. K-means algorithm is applied to $ \mathbf{U}_k $, producing clusters $ \mathcal{C} = \{C_{1},C_{2},\dots, C_k \} $. For cluster $ C_p $ with $ m $ elements, the elements are represented as
\begin{equation}
C_p = \underbrace{\{h, \cdots, j\}}_{m \ \text{elements in total}}.
\end{equation}
\subsubsection{Cluster Membership Complement}
Let the initial cluster set be $\mathcal{C}^{(0)}=\{C_{1}^{(0)},C_{2}^{(0)},\dots,C_{k}^{(0)}\}$, with a prescribed minimum cluster size $b\in\mathbb{N}_{+}$\footnote{The value of $b$ is related to the number of spatial dimensions, this is because relative coordinates need anchors/public nodes to be transformed and merged, usually $b=4$ in 3D space. \cite{swarm}\cite{Denied}}. The inter‑cluster range in the $t$-th iteration is defined as 
\begin{equation}
\delta(C_{p}^{(t)},C_{q}^{(t)})=\min_{u\in C_{p}^{(t)}\ v\in C_{q}^{(t)}}d_{uv},\qquad p\neq q,
\end{equation}
and the adjacent cluster of $C_{p}^{(t)}$ is defined as 
\begin{equation}
\operatorname{Adj}\bigl(C_{p}^{(t)}\bigr)=\arg\min_{C_{q}^{(t)}\in\mathcal{C}^{(t)}\setminus\{C_{p}^{(t)}\}}\delta\bigl(C_{p}^{(t)},C_{q}^{(t)}\bigr).
\end{equation}
For the initial cluster set $\mathcal{C}^{(0)}$, the size of each cluster, i.e. the number of elements in each cluster $|C_{p}^{(0)}|$ is computed, where $|\cdot|$ denotes the number of elements in the set. Let 
\begin{equation}
\mathcal{S}^{(0)}=\bigl\{ C_{p}^{(0)}\in\mathcal{C}^{(0)}\;|\;|C_{p}^{(0)}|<b \bigr\}
\end{equation}
be the set of undersized clusters. In each iteration, when $\mathcal{S}^{(t)}\neq\varnothing$, the following merging operation is implemented for each $C_{p}^{(t)}\in\mathcal{S}^{(t)}$
\begin{equation}
C_{q}^{\ast}=\arg\min_{C_{q}^{(t)}\in\mathcal{C}^{(t)}\setminus\{C_{p}^{(t)}\}}\delta(C_{p}^{(t)},C_{q}^{(t)}),
\end{equation}
 and the cluster set is updated by 
\begin{equation}
\mathcal{C}^{(t+1)}\leftarrow \bigl(\mathcal{C}^{(t)}\setminus\{C_{p}^{(t)},C_{q}^{\ast}\}\bigr)\cup\{C_{p}^{(t)}\cup C_{q}^{\ast}\}. 
\end{equation}
 After each merge, $\mathcal{S}^{(t+1)}$ is recomputed, and this process continues until $|C_{i}^{(t)}|\ge b$ for all $C_{i}^{(t)}\in\mathcal{C}^{(t)}$. Since $\delta(\cdot,\cdot)$ is symmetric and each merge strictly reduces the number of clusters, the process will terminate in a finite number of steps, yielding the final cluster set $\mathcal{C}$ that satisfies the minimum cluster size constraint.
%
%
%

\subsection{Intra-cluster Localization}\label{intr}

\subsubsection{Public Node Selection}\label{sel}

For each UAV $ u \in C_p $, the minimum range to $ \operatorname{Adj}(C_p) $ is calculated as
\begin{equation}
d_u^{\,\operatorname{adj}} = \min_{v \in \operatorname{Adj}(C_p)} \hat{d}_{uv}.
\end{equation}
The two nodes corresponding to the smallest ranges are selected as $ \mathcal{A}_p = \{a_p^{(1)}, a_p^{(2)}\} \subset C_p $. 

Similarly, within $ \operatorname{Adj}(C_p) $, $ \mathcal{A}_q = \{a_q^{(1)}, a_q^{(2)}\} \subset \operatorname{Adj}(C_p) $ is obtained, resulting in four public nodes: 
\begin{equation}
\mathcal{P}_{\text{pub}} = \mathcal{A}_p \cup \mathcal{A}_q, \qquad |\mathcal{P}_{\text{pub}}| = 4,
\end{equation}
The external anchor nodes are then incorporated into their respective clusters: 
\begin{equation}
C_p \leftarrow C_p \cup \mathcal{A}_q, \qquad \operatorname{Adj}(C_p) \leftarrow \operatorname{Adj}(C_p) \cup \mathcal{A}_p, 
\end{equation}

\subsubsection{Matrix Complement}

After clustering,  for a cluster $C_{p}$ containing $m$ UAVs, the range matrix can be constructed for cluster $C_{p}$ as 
\begin{equation}
\mathbf{D}_{C_{p}}=\left( \begin{matrix} 

0 &\hat{d}_{h,i}& \cdots & \hat{d}_{h,j} \\ 
\hat{d}_{i,h} &0& \cdots & \vdots \\ 

\vdots&\vdots & \ddots & \vdots \\ 

\hat{d}_{j,h} &\cdots& \cdots & 0 \\ 

\end{matrix} \right).
\end{equation}
where $\mathbf{D}_{C_{p}}\in \mathbb{R}^{m \times m}$, its diagonal element is 0. However, the UAVs inside a cluster do not necessarily form a fully connected ranging graph. Consequently, the corresponding observed range matrix $\mathbf{D}_{C_p} \in \mathbb{R}^{m \times m}$ still contains missing entries and is only specified at indices $(i, j) \in \Omega$, where $\Omega$ denotes the set of index pairs with available ranging measurements. The matrix‑completion technique proposed in \cite{MATCOM} is adopted to reconstruct the complete set of inter-agent ranges. The method approximates the incomplete matrix by a low‑rank factorization $\mathbf{D}\approx\mathbf{U}\mathbf{V}^{\top}$ and refines the factors via Alternating Least Squares (ALS) until a filled matrix $\hat{\mathbf{D}}$ is obtained. The optimisation problem is formulated as 
\begin{equation}
\min_{\mathbf{U},\mathbf{V}}
\sum_{(i,j)\in\Omega}\!\bigl(d_{ij}-[\mathbf{U}\mathbf{V}^{\top}]_{ij}\bigr)^{2}
+\lambda\bigl(\|\mathbf{U}\|_{F}^{2}+\|\mathbf{V}\|_{F}^{2}\bigr),
\end{equation}
where the $d_{ij}$ is the element of $\mathbf{D}$, and the first term penalises the reconstruction error on observed entries, and the second term is the Frobenius‑norm regulariser, weighted by $\lambda>0$, prevents over‑fitting and stabilises the estimates of unobserved elements. 

The optimization problem described above is bi-variable and non-convex (it is non-convex concerning the joint optimization of $ \mathbf{U} $ and $ \mathbf{V} $, but convex for each variable individually). Therefore, by fixing $ \mathbf{V} $, the problem becomes a convex quadratic optimization problem concerning $ \mathbf{U} $. Fixing $ \mathbf{U} $, the problem becomes a convex quadratic optimization problem concerning $ \mathbf{V} $. This can be solved using ALS. Initially, $ \mathbf{U} $ and $ \mathbf{V} $ are randomly initialized. Fixing $ \mathbf{V} $, each row $ \mathbf{u}_i $ is solved by
\begin{equation}\label{u1}
\mathbf{u}_i = \left( \sum_{j \in \Omega_i} \mathbf{v}_j \mathbf{v}_j^{\top} + \lambda I \right)^{-1} \left( \sum_{j \in \Omega_i} \mathbf{D}_{ij} \mathbf{v}_j \right) .
\end{equation}
Similarly, by fixing $ \mathbf{U} $, each column $ \mathbf{v}_j $ is solved as: 
\begin{equation}\label{u2}
\mathbf{v}_j = \left( \sum_{i \in \Omega_j} \mathbf{u}_i \mathbf{u}_i^{\top} + \lambda I \right)^{-1} \left( \sum_{i \in \Omega_j} \mathbf{D}_{ij} \mathbf{u}_i \right). 
\end{equation}
 This process is iterated until convergence, yielding the complete matrix

%
\begin{equation}\label{D_com}
\hat{\mathbf{D}}=\mathbf{U}^{{\left(K_{\text{max}}\right)}}\mathbf{V}^{\left(K_{\text{max}}\right)\top},
\vspace{-8pt}
\end{equation}
where $K_{\text{max}}$ denotes the preset maximum number of iterations. $\mathbf{D}_{C_{p}}$ can be complemented by \eqref{u1}, \eqref{u2} and \eqref{D_com} after $K_{\text{max}}$ iterations. The process provides gap‑free range information for subsequent MDS‑based coordinate recovery.
\subsubsection{Relitive Localization Based on MDS}\label{MDS}

%
%
%
%
In MDS, the pairwise range information between nodes is used to construct constraints, and SVD is applied to find an $n$-dimensional coordinate representation that satisfies these range constraints. For the cluster $C_{p}$ consisting of $m$ UAVs, to eliminate the translational effects of the coordinates, the centering matrix is introduced that removes the translation impact, ensuring that the coordinates are symmetric about the origin
\begin{equation}
\mathbf{J}_{C_{p}} = \mathbf{I} - \frac{1}{m} \mathbf{1} \mathbf{1}^{\top} ,
\end{equation}
 where $\mathbf{1}$ is a vector of ones, which's length is $m$. By converting the squared range matrix,  the positive-semidefinite $\mathbf{B}_{C_{p}}$ can be expressed as
\begin{equation}
\mathbf{B}_{C_{p}} = -\frac{1}{2} \mathbf{J}_{C_{p}} \mathbf{D}_{C_{p}}^{\odot 2} \mathbf{J}_{C_{p}},
\end{equation}
where $\{\cdot\}^{\odot 2}$ denotes the element-wise squaring operation, and $\mathbf{D}_{C_{p}}^{\odot 2}$ represents the squared pairwise ranges between UAVs in the cluster. Next, Singular value decomposition (SVD) is implemented on $\mathbf{B}_{C_{p}}$, yielding
\begin{equation}
\mathbf{B}_{C_{p}} = \mathbf{U}_{C_{p}} \boldsymbol{\varLambda}_{C_{p}} \mathbf{V}_{C_{p}}^{\top} ,
\end{equation}
where $\mathbf{U}_{C_{p}}$ is the matrix of eigenvectors, each column representing an eigenvector, and $\boldsymbol{\varLambda}_{C_{p}}$ is the diagonal matrix of eigenvalues. The first three eigenvectors are selected corresponding to the smallest eigenvalues to form the matrix $\mathbf{U}^{\prime}_{C_{p}} \in \mathbb{R}^{m \times 3}$ and the diagonal matrix $\boldsymbol{\varLambda}^{\prime}_{C_{p}} \in \mathbb{R}^{3 \times 3}$. Finally, the 3D coordinate matrix $\mathbf{P}^{C_p}_{\text{rel}}$ is obtained for the UAVs in cluster $C_{p}$ as 
\begin{equation}
\mathbf{P}^{C_p}_{\text{rel}} = \mathbf{U}^{\prime}_{C_{p}} \boldsymbol{\varLambda}^{\prime}_{C_{p}} .
\end{equation}

\subsection{Global Map Compositing}
\subsubsection{Procrustes Analysis}

After public node selection in the Section \ref{sel}, adjacent clusters typically share a subset of public nodes. Within each cluster, relative localization has been completed using MDS combined with matrix completion. Local coordinate maps must be merged to obtain a global coordinate system for the UAV swarm. Cluster $ C_p $ consisting of $ m $ UAVs with relative positions is denoted by $ \mathbf{P}^{C_p}_{\text{rel}} = [ \mathbf{p}^{C_p}_1, \mathbf{p}^{C_p}_2, \cdots, \mathbf{p}^{C_p}_m ]^{\top} $. To merge cluster $ C_p $ into the reference cluster $ C_q $, the coordinates of four public nodes of $C_{p}$ and $C_{q}$ are identified as $ \mathbf{P}^{C_p}_{\text{pub}} \in \mathbb{R}^{4 \times 3} $ and $ \mathbf{P}^{C_q}_{\text{pub}} \in \mathbb{R}^{4 \times 3} $. The transformation from $ C_p $ to $ C_q $'s coordinate system is accomplished via Procrustes analysis, which applies rotation and translation to align the coordinate frames. As illustrated in \cite{swarm}, the transformation is given by
\begin{equation}
\mathbf{P}^{C_{q} \leftarrow C_{p}}_{\text{rel}} = \mathbf{R} \mathbf{P}^{C_{p} \top}_{\text{rel}} + \mathbf{t} \mathbf{1}^\top_{C_p},
\end{equation}
where $ \mathbf{R}  \in \mathrm{SO}(3)  $\footnote{$\mathrm{SO}(3)$ refers to the Special Orthogonal Group in 3D, which is the set of all $3\times 3$ real rotation matrices that preserve the geometry of 3D space.
} is the rotation matrix and $ \mathbf{t} \in \mathbb{R}^{3 \times 1} $ is the translation vector. The optimal transformation can be obtained by solving the following MMSE problem
\begin{equation}\label{pro}
\{\mathbf{R}, \mathbf{t} \} = \arg \min_{\mathbf{R}', \mathbf{t}'} \sum_{i=1}^{4} \| \mathbf{R}' \mathbf{p}^{C_p}_i + \mathbf{t}' - \mathbf{p}^{C_q}_i \|^2_2.
\end{equation}
To solve \eqref{pro}, the translation bias should be removed firstly, both coordinate sets are mean-centered
\begin{equation}\label{nomean}
\bar{\mathbf{P}}^{C_{p}}_{\text{pub}} = {\mathbf{P}}^{C_{p}}_{\text{pub}} - \mathbf{1}_{\text{pub}} \boldsymbol{\mu}^{C_{p} \top}_{\text{pub}},\quad 
\bar{\mathbf{P}}^{C_{q}}_{\text{pub}} = {\mathbf{P}}^{C_{q}}_{\text{pub}} - \mathbf{1}_{\text{pub}} \boldsymbol{\mu}^{C_{q} \top}_{\text{pub}},
\end{equation}
where $ \boldsymbol{\mu}^{C_{p}}_{\text{pub}} = \frac{1}{4} \sum_{i=1}^4 \mathbf{p}^{C_p}_i $ and $ \boldsymbol{\mu}^{C_{q}}_{\text{pub}} = \frac{1}{4} \sum_{i=1}^4 \mathbf{p}^{C_q}_i $. SVD is implemented to the cross-covariance matrix
\begin{equation}\label{dep}
\bar{\mathbf{P}}^{C_{q} \top}_{\text{pub}} \bar{\mathbf{P}}^{C_{p}}_{\text{pub}} = \mathbf{U} \boldsymbol{\varLambda} \mathbf{V}^\top.
\end{equation}
The rotation matrix and translation vector can be expressed as 
\begin{equation}\label{result}
\mathbf{R} = \mathbf{U} \mathbf{V}^\top,\quad 
\mathbf{t} = \boldsymbol{\mu}^{C_{q}}_{\text{pub}} - \mathbf{R} \boldsymbol{\mu}^{C_{p}}_{\text{pub}}.
\end{equation}
The final merged coordinates under the $ C_q $ reference coordinate frame are
\begin{equation}\label{trans}
\mathbf{P}^{C_{q} \leftarrow C_{p}}_{\text{rel}} = 
\begin{cases} 
\mathbf{R} \mathbf{p}^{C_p}_i + \mathbf{t}, & i \in C_p \setminus \{C_{p}\cap C_{q}\} \\ 
\mathbf{p}^{C_q}_i, & i \in C_q \setminus \{C_{p}\cap C_{q}\} \\ 
\frac{1}{2} (\mathbf{R} \mathbf{p}^{C_p}_i + \mathbf{t} + \mathbf{p}^{C_q}_i), & i \in C_{p} \cap C_{q}
\end{cases}.
\end{equation}
This completes the transformation of all UAV coordinates in cluster $ C_p $ into the coordinate frame of cluster $ C_q $.  

\begin{figure}[!t]
\centering
\subfloat[]{\includegraphics[width=3.2in]{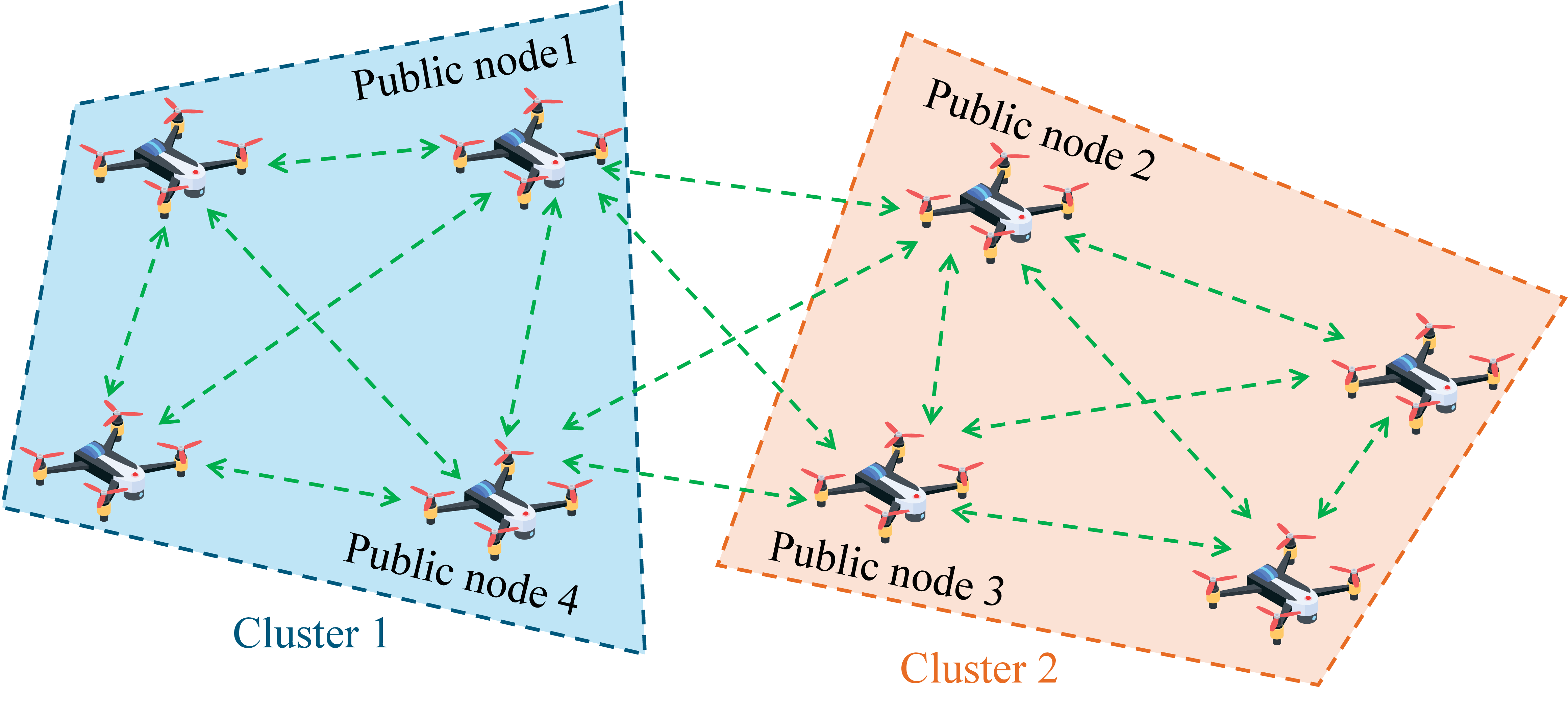}%
\label{Merge_a}}
\hfil
\vspace{-12pt}
\subfloat[]{\includegraphics[width=3.2in]{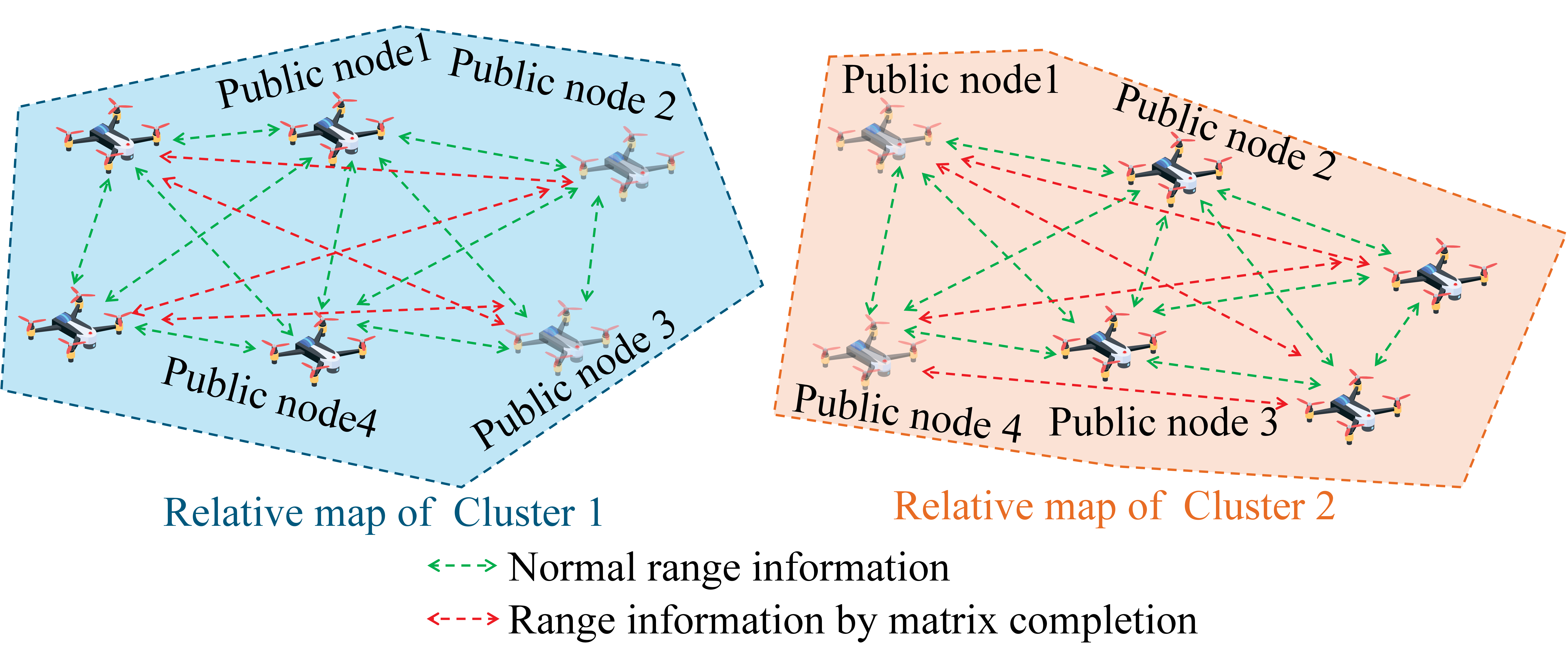}%
\label{Merge_e}}
\hfil
\vspace{-12pt}
\subfloat[]{\includegraphics[width=3.2in]{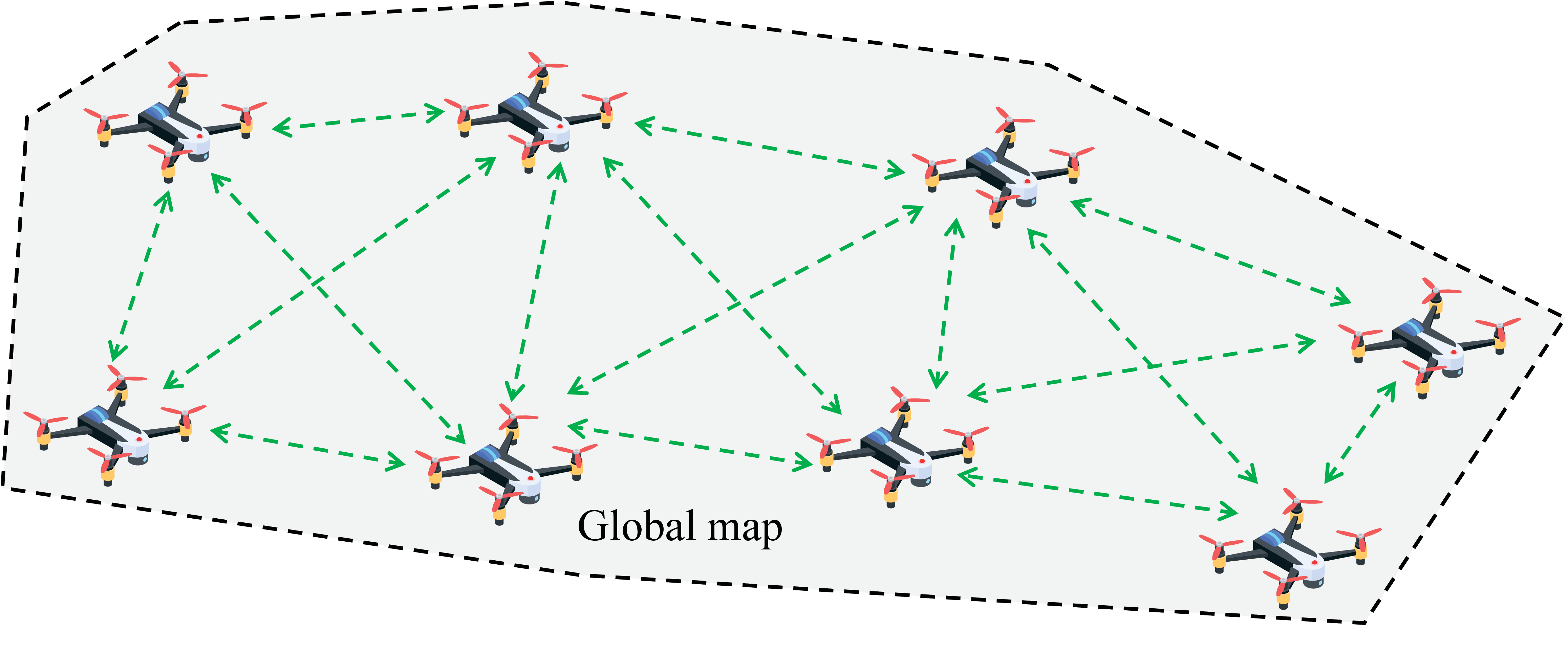}%
\label{Merge_d}}
\caption{(a) The distribution of Cluster 1 and Cluster 2 membership. (b) The construction of relative maps of cluster 1 and cluster 2. (c) The construction of the global map.}
\label{Merge}
\vspace{-10pt} 
\end{figure}

\subsubsection{Cluster Merge}

Let the initial cluster set after clustering be $ \mathcal{C}^{(0)} = \{ C_1^{(0)},C_2^{(0)} \dots, C_k^{(0)} \} $. The reference cluster is chosen as 
\begin{equation}
C_r^{(0)} = \arg\max_{1 \le p \le k} |C_p^{(0)}|, \quad \mathcal{T}^{(0)} = \mathcal{C}^{(0)} \setminus \{ C_r^{(0)} \}, 
\end{equation}
where $ \mathcal{T}^{(t)} $ denotes the set of clusters that have not yet been merged into the global map during the $ t $-th iteration. At the $ t $-th iteration ($ t \geq 0 $), the cluster with a global coordinate is denoted as $ C_r^{(t)} $, and its neighboring clusters are defined as
\begin{equation}
C_a^{(t)} = \operatorname{Adj}_{C \in \mathcal{T}^{(t)}} \left( C_r^{(t)} \right), 
\end{equation}
i.e., the cluster with the smallest inter-cluster range to $ C_r^{(t)} $. The smallest four elements from the range matrix are selected to form the set $ \{C_r^{(t)} \cup C_a^{(t)} \}$. Let the local coordinates obtained by MDS be $ \mathbf{P}_{\text{rel}}^{(t)C_{r}} \in \mathbb{R}^{3 \times |C_r^{(t)}|} $ and $ \mathbf{P}_{\text{rel}}^{(t)C_{a}} \in \mathbb{R}^{3 \times |C_a^{(t)}|} $. The rotation matrix $ \mathbf{R}^{(t)}  $ and translation vector $ \mathbf{t}^{(t)}  $ is obtained by \eqref{nomean}, \eqref{dep} and \eqref{result}, so that ${\mathbf{P}}_{\text{rel}}^{(t)C_{r}\leftarrow C_{a}}$ is obtained by \eqref{trans}.
 After merging, the new global map and point set is obtained as
\begin{equation}
\hat{\mathbf{P}}^{(t+1)} = \left[ \mathbf{P}_{\text{rel}}^{(t)C_{r}}, \; {\mathbf{P}}_{\text{rel}}^{(t)C_{r}\leftarrow C_{a}} \right], \quad C_r^{(t+1)} = C_r^{(t)} \cup C_a^{(t)}. 
\end{equation}
 Then, update $ \mathcal{T}^{(t+1)} = \mathcal{T}^{(t)} \setminus \{ C_a^{(t)} \} $, and the process ends when $ \mathcal{T}^{(t)} = \varnothing $. The whole process of merging clusters is shown in Fig.~\ref{Merge}. Since each iteration reduces the number of clusters by one, the algorithm converges after at most $ k-1 $ steps, outputting the UAV positions in a unified coordinate system.

\section{Case study and Numerical Results}
\subsection{OTFS Signal Model}

OTFS signals are modulated in the delay-Doppler (DD) domain, with data symbols mapped onto the delay-Doppler resource element (DDRE) of $ M \times N $. The transmitted signal at each grid point is represented as $ \mathbf{X}[k,l] $, where $ k $ and $ l $ are the Doppler and delay indices ($ k = 0, 1, \dots, N-1 $, $ l = 0, 1, \dots, M-1 $). The grid resolutions in delay and Doppler are $ \frac{1}{M \Delta f} $ and $ \frac{1}{NT} $, corresponding to ranges of $ T $ and $ -\Delta f / 2 $ to $ \Delta f / 2 $, respectively.

In practical systems, delay and Doppler shifts are often not integer multiples of the DDRE grid step sizes, leading to fractional components, known as fractional effects. These cause energy to diffuse within the DDRE. The delay and Doppler values of the $ p $-th path are defined as $ \tau_p $ and $ \nu_p $, where $ \nu_p = \frac{k_{\nu_p} \Delta f}{N} = \frac{(k_p + \kappa_p) \Delta f}{N} $ and $ \tau_p = \frac{l_{\tau_p} T}{M} = \frac{(l_p + \iota_p) T}{M} $, with corresponding indices $ k_{\nu_p} $ and $ l_{\tau_p} $, respectively. Here, $ k_{\nu_p} = k_p + \kappa_p $ and $ l_{\tau_p} = l_p + \iota_p $, where $ \kappa_p $ and $ \iota_p $ are fractional terms within $ [-0.5, 0.5] $. The input-output relationship in the DD domain is expressed as

\begin{equation}\label{YDDHDD}
\begin{aligned} 
\mathbf{Y}[k,l] &= \mathbf{H}[k,l] \circledast \mathbf{X}[k,l] + \mathbf{Z}[k,l] \\ 
&= \sum_{k^\prime=0}^{N-1}\sum^{M-1}_{l^\prime=0} \mathbf{X}[k^\prime,l^\prime] \mathbf{H}[[k - k^\prime]_N, [l - l^\prime]_M] + \mathbf{Z}[k, l], 
\end{aligned}
\end{equation}
where $\circledast$ denotes the two-dimensional circular convolution, and $\mathbf{Z} \in \mathbb{C}^{M \times N}$ represents $i.i.d$ complex Gaussian noise. $\mathbb{C}^{M \times N}$ denotes the set of all $M \times N$ matrices with complex entries. $\alpha_{p}$ denotes the channel gain of the $p$-th path, which is related to the reflection and scattering characteristics of the reflectors in the scene. $\mathbf{H}$ can be rewritten as the sum of responses from different paths on the DDRE
\begin{equation}\label{hdddd}
\mathbf{H}[k,l] = \sum_{p=1}^{P} \alpha_p \frac{\mathbf{v}_{\tau_p,l}^{T} \mathbf{W}_{\text{\text{in}}} \mathbf{v}_{\nu_p,k}}{MN} \exp\left(-j \frac{2\pi k_{\nu_p} l_{\tau_p}}{MN}\right), 
\end{equation}
where
\begin{equation}\label{VV}
\begin{aligned} 
\begin{cases} 
&\mathbf{v}_{\nu_p, k}(n) = \exp\left(-j \frac{2\pi n (k - k_{\nu_p})}{N}\right), \quad n = 0, 1, \dots, N-1, \\ 
&\mathbf{v}_{\tau_p, l}(m) = \exp\left(j \frac{2\pi m (l - l_{\tau_p})}{M}\right), \quad m = 0, 1, \dots, M-1, 
\end{cases} .
\end{aligned} 
\end{equation}
 $\mathbf{v}_{\nu_p, k} \in \mathbb{C}^{N \times 1}$, $\mathbf{v}_{\tau_p, l} \in \mathbb{C}^{M \times 1}$, and $\mathbf{W}_{\text{\text{in}}} = \mathbf{1}^{M \times N}$.


Range estimation can be accomplished by delay estimation. The embedded pilot structure of OTFS in \cite{Embeded} is introduced. Channel estimation method in \cite{Estimation} is applied to obtain the delay estimation $\hat{\tau}_{p}$, the Doppler estimation $\hat{\nu}_{p}$, and the channel gain estimation $\hat{\alpha}_{p}$, and the delay estimation between two UAVs can be transformed into the range estimation between two UAVs, i.e. $\hat{d}_{ij}=c\hat{\tau}_{ij}$, to complete the subsequent relative swarm localization.

%

\begin{figure}[htbp]
\centering
\includegraphics[width=0.3\textwidth]{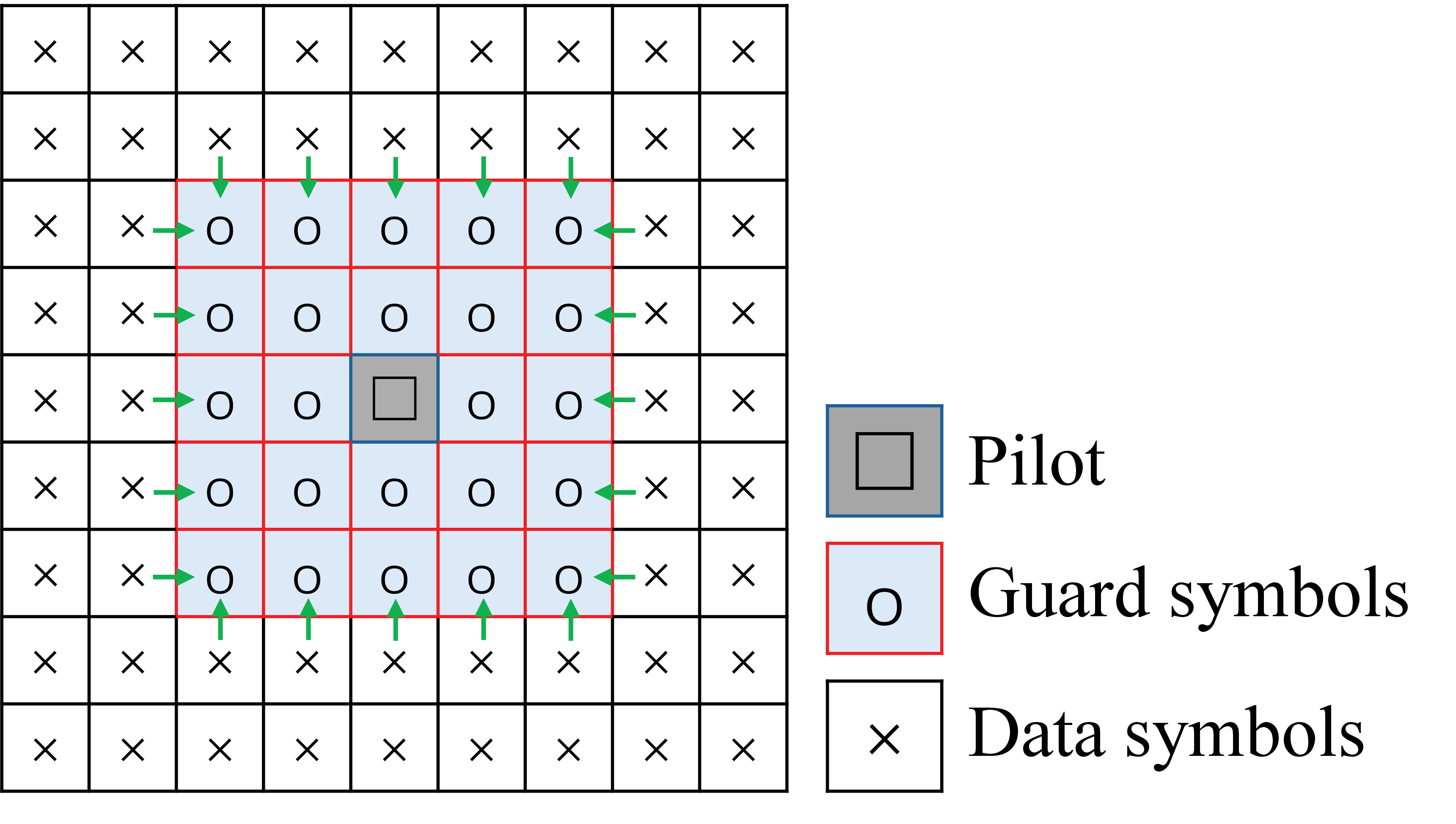}
\caption{The embeded pilot structure of OTFS signals. At the receiver, the channel estimation will be implemented on the guard symbols, where the data symbols' energy will diffuse because of the fractional effect.}
\label{Pilot}
\vspace{-20pt}

\end{figure}

\subsection{Simulation Setup}
In the numerical simulation, $L=50$ UAVs are randomly distributed in a 1000m $\times$ 1000m $\times$ 1000m 3D space. To evaluate the localization performance under varying connectivity conditions, the range measurement retention ratio is introduced to assess the loss of ranging information. Initially, the UAV network is fully connected, and ranging information is progressively removed based on the retention ratio. 

To validate the proposed algorithm, classical network localization methods such as MDS-MAP \cite{shang2003localization} and MDS-MAP(P) \cite{MDSMAPP} are used for comparison. The number of clusters in the spectral clustering method is fixed at $ k = 6 $, providing a benchmark for evaluating adaptive clustering effectiveness. After performing relative localization on the UAVs, the absolute positions of four UAVs are selected and aligned with the actual positions of the UAV network. Localization accuracy is measured using the root mean square error $(\text{RMSE} = \sqrt{\frac{1}{L} \sum_{i=1}^{L} ||\mathbf{p}_i - \hat{\mathbf{p}}_i||^2})$: 


The case study uses OTFS signal channel estimation to obtain range measurements and perform UAV swarm localization. The DDRE parameters are set to $ M = 32 $, $ N = 32 $, $ \Delta f = 552.3 $ kHz, and $ f_c = 5.1 $ GHz. The pilot signal-to-noise ratio is set to 25 dB, with data symbol power adjusted between -20 and 0 dB relative to pilot power. The channel contains only one LoS path. Range measurements are transformed by $ \hat{d} = c\hat{\tau} $. Bit error rate (BER) measures communication performance, and RMSE measures localization accuracy.


\subsection{Results Discussion}
\subsubsection{Localization Performance}
\begin{figure}[htbp]
\centering
\includegraphics[width=0.4\textwidth]{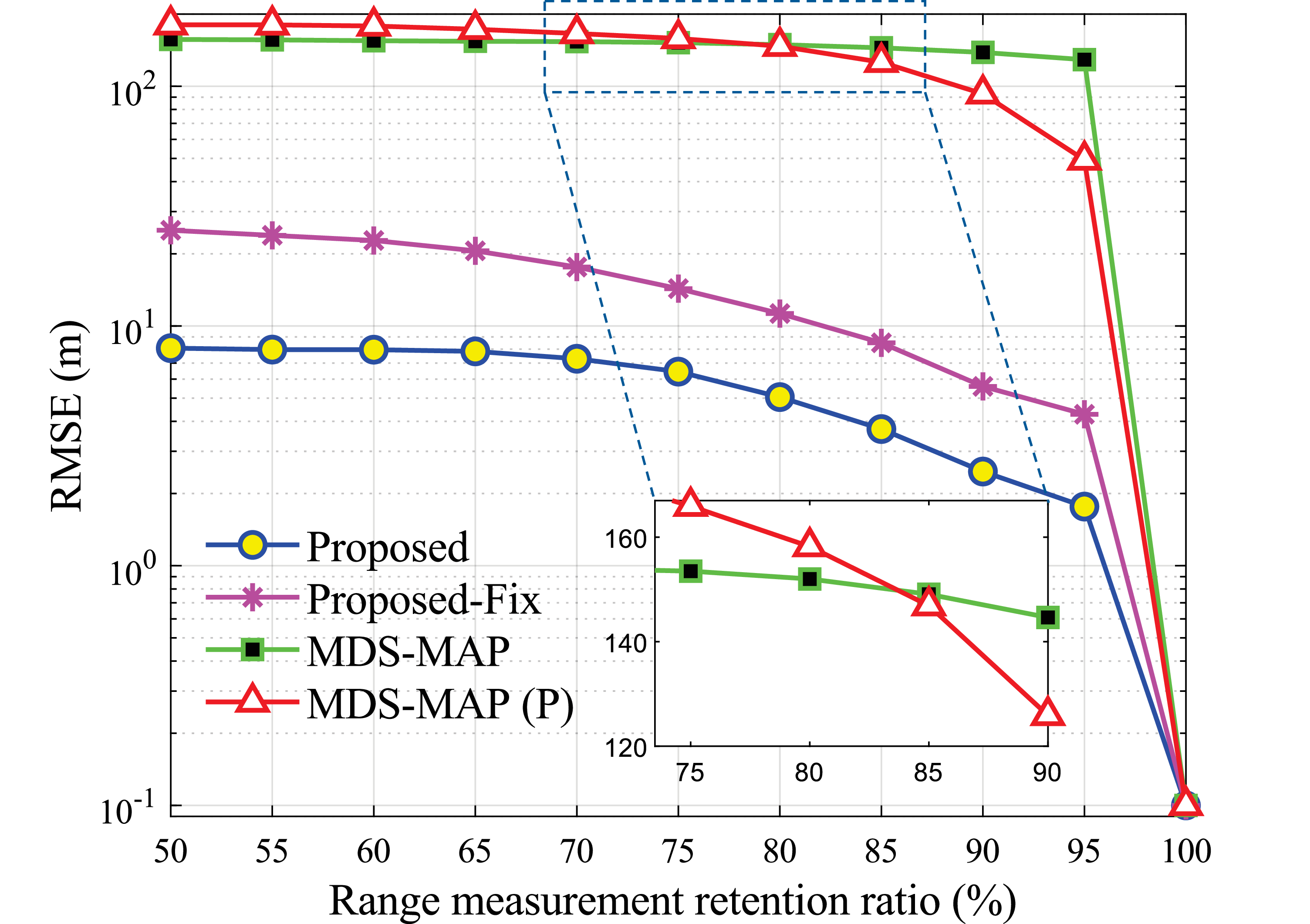}
\caption{The localization accuracy versus the range measurement retention ratio.} 
\label{fig:Results_1}
\vspace{-15pt} 
\end{figure}

In Fig.~\ref{fig:Results_1}, both the proposed method and proposed-Fix outperform MDS-MAP and MDS-MAP (P) in localization accuracy. The matrix completion in the proposed method compensates for missing observations, while traditional methods rely on less accurate multi-hop ranges. The overall ranging error of proposed-Fix is higher than the proposed method due to the latter’s use of node similarities and adaptive clustering, which ensures fully connected clusters and reduces intra-cluster localization errors. As the range measurement retention ratio increases, the network approaches full connectivity, and RMSE decreases for all algorithms when no observations are missing, converging with real network localization behavior. When 50\%-80\% of the range information is missing, MDS-MAP (P) performs worse than MDS-MAP due to error accumulation during cluster filling and map merging. With less missing data, MDS-MAP (P) performs better than MDS-MAP due to fewer errors within clusters.

\begin{figure}[!t]
\centering
\subfloat[]{\includegraphics[width=1.7in]{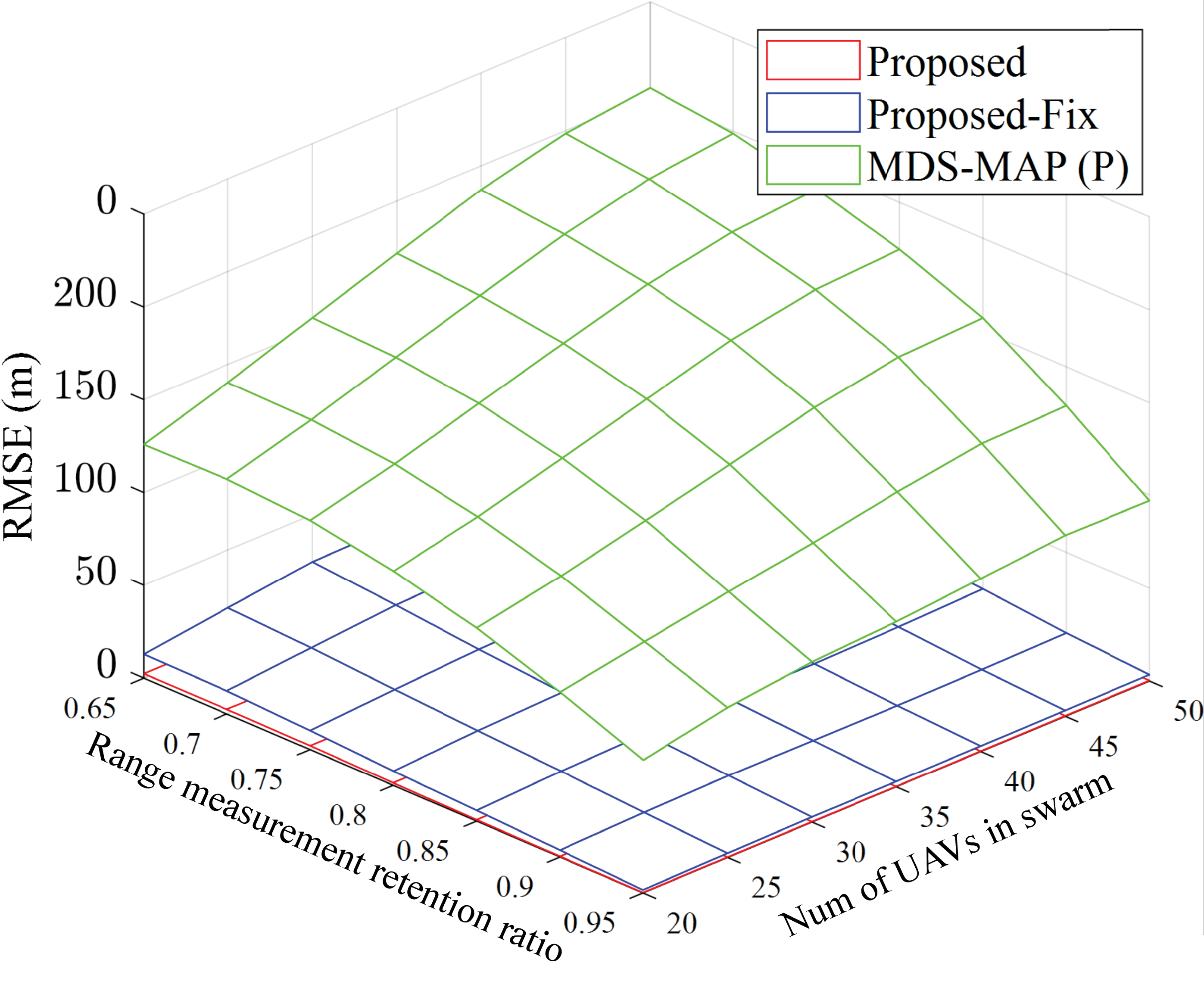}%
\label{Results_3}}
\hfil
\subfloat[]{\includegraphics[width=1.7in]{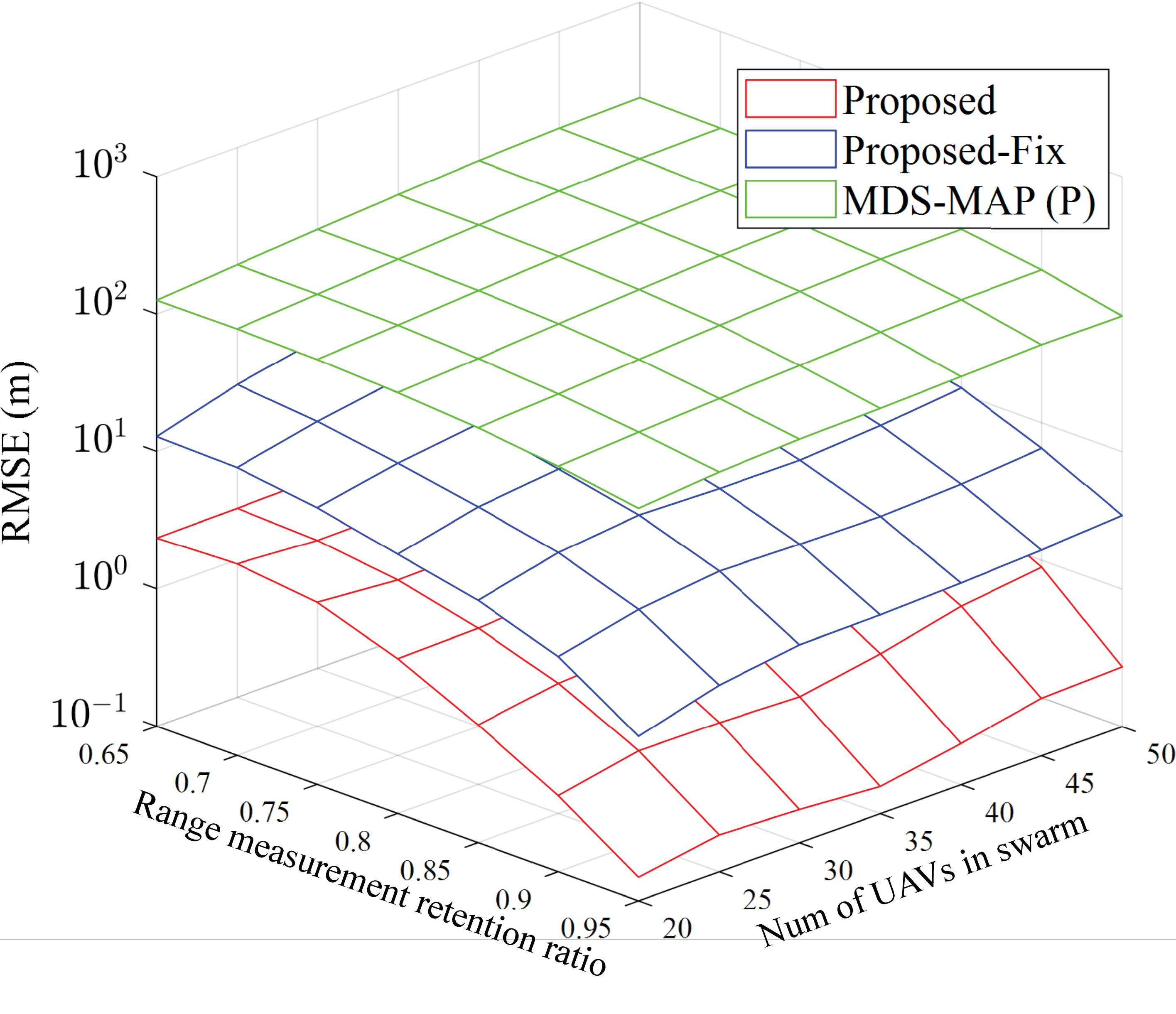}%
\label{Results_4}}
\hfil
\vspace{-8pt}
\caption{The localization accuracy versus the range measurement retention ratio and number of UAVs in the swarm.}
\label{Results34}
\vspace{-20pt} 
\end{figure}

The localization performance of different methods under varying range measurement retention ratios and the number of UAVs in the swarm is shown in Fig.~\ref{Results34}. Overall, the proposed method and proposed-Fix outperform MDS-MAP (P) under all conditions. The proposed method ensures minimal loss of range information within each cluster, resulting in fewer errors caused by compensating for missing observations, and thus provides better localization performance compared to the proposed-Fix. Additionally, as the number of UAVs increases and the range measurement retention ratio decreases, the localization performance of all methods declines, which aligns with real-world behavior.

\subsubsection{ISAC Performance}
\begin{figure}[htbp]
\centering
\includegraphics[width=0.4\textwidth]{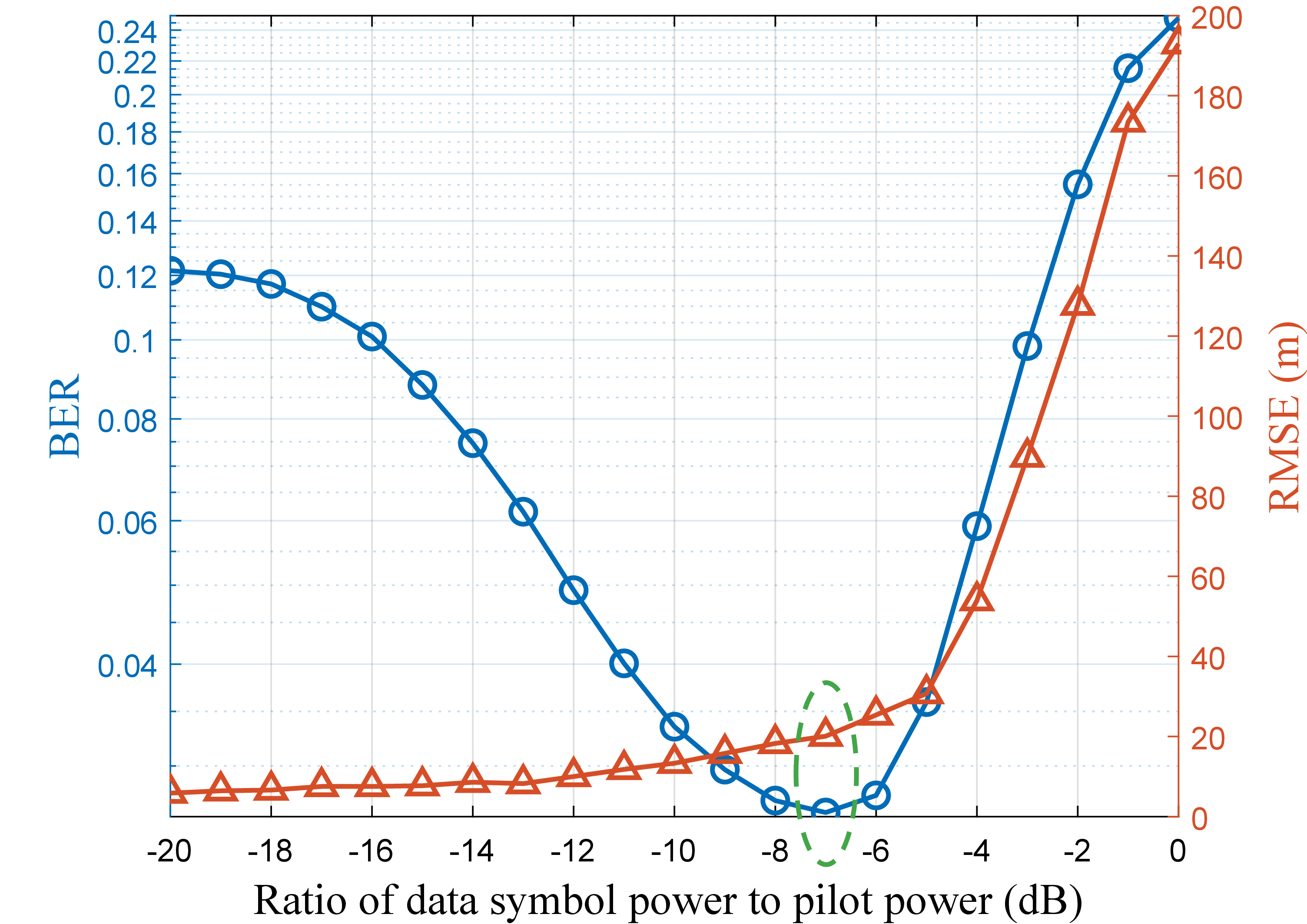}
\caption{System communication and localization performance versus ratio of data symbol power to pilot power.} 
\label{fig:Results_2}
\vspace{-20pt} 
\end{figure}

In Fig.~\ref{fig:Results_2}, the system's BER decreases and rises as data symbol power increases. Increasing symbol power boosts SNR, reducing BER. However, fractional delay and Doppler effects on the pilot block introduce contamination, affecting delay estimation and increasing localization RMSE. As data symbol's power rises, poor channel estimation exacerbates the issue, causing BER to spike. When the ratio of data symbol power to pilot power exceeds -7 dB, both BER and RMSE increase rapidly, significantly degrading communication and localization performance. This underscores the impact of parameter adjustments in the ISAC system, highlighting future work on optimizing the trade-offs between communication and localization.

\section{Conclusion}

The paper proposes a hierarchical cooperative localization framework for UAV swarm localization in GNSS-denied environments, addressing missing inter-UAV range information in large swarms. Communication signal channel estimation forms a ranging graph partitioned using spectral clustering. MDS and matrix completion perform relative localization within clusters, while inter-cluster fusion of public nodes creates a global map. OTFS signals build an ISAC system. Experimental results show improved localization accuracy and highlight the relationship between communication performance and localization precision, supporting the development of high-performance ISAC UAV localization systems.

\bibliographystyle{IEEEtran}
\bibliography{my_ref}
\end{document}